\documentclass{PoS}

\usepackage{graphicx}
\usepackage{amssymb}
\usepackage{amsmath}

\newcommand{\tr}{\mathrm{tr}}
\newcommand{\Retr}{\mathrm{Re \, tr}}

\title{Rotating lattice}

\ShortTitle{Rotating lattice}

\author{\speaker{Arata~Yamamoto}\\
Theoretical Research Division, Nishina Center, RIKEN, Saitama 351-0198, Japan\\
E-mail: \email{arayamamoto@riken.jp}}

\author{Yuji~Hirono\\
Theoretical Research Division, Nishina Center, RIKEN, Saitama 351-0198, Japan\\
Department of Physics, The University of Tokyo, Tokyo 113-0033, Japan\\
Department of Physics, Sophia University, Tokyo 102-8554, Japan}

\abstract{
We present the lattice QCD formulation in rotating frames.
We start with the continuum QCD action in rotating frames, and then discretize it on the lattice.
For the first test of the formulation, we calculate angular momentum in the quenched Monte Carlo simulation, and confirm that the formulation works successfully.
}

\FullConference{The XXXI International Symposium on Lattice Field Theory\\
July 29-August 3, 2013\\
Mainz, Germany}

\begin{document}

\section{Introduction}
So far, people have introduced several extreme conditions or external environments to lattice QCD, for example, a finite temperature, a finite density, and an external electromagnetic field.
In this study, we introduce another extreme condition, that is, a ``rotation''.
There are many rotations in QCD.
The first example is the ultrarelativistic heavy-ion collision at Relativistic Heavy Ion Collider (RHIC) or Large Hadron Collider (LHC).
In a non-central collision of heavy ions, the fragments have large angular momenta around the central axis of the collision.
The created quark-gluon plasma and the created hadrons rotate rapidly.
The second example is the inner core of a rapidly rotating compact star.
The core of a compact star is a high-density QCD matter like a color superconductor, and it rotates because of the rotation of the star.
Also in low-energy nuclear physics, rotational modes or high-spin states of nuclei are interesting theoretical topics, and they are produced in nuclear experimental facilities.

\begin{figure}[h]
\begin{center}
\includegraphics[scale=0.5]{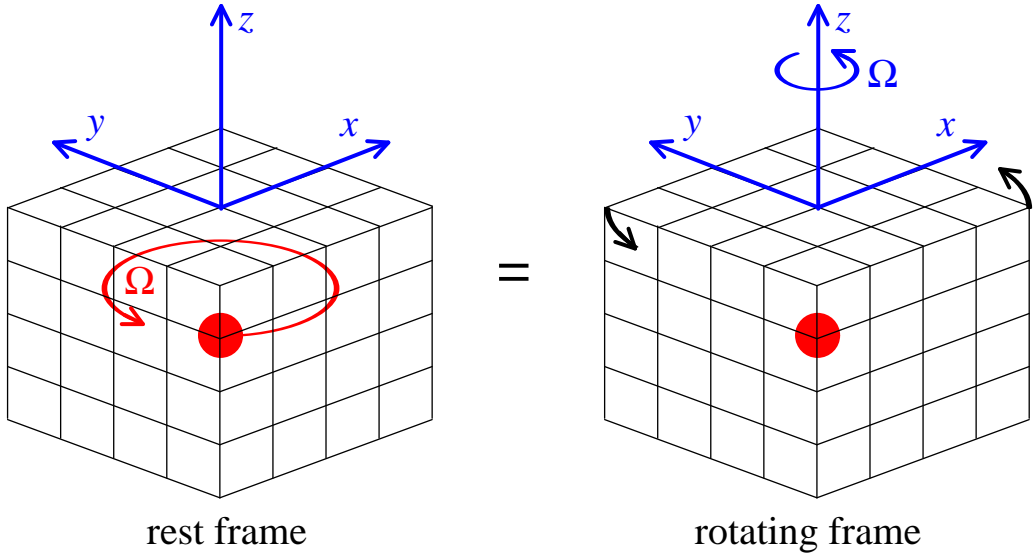}
\caption{\label{fig1}
Rotating lattice.
}
\end{center}
\end{figure}

However, the lattice QCD simulation of a rotating matter is quite difficult in the straightforward way.
Rotation is characterized by the circulation of velocity field.
In a rotating matter, particles flow with finite velocity.
Such a state is not in equilibrium.
Since the lattice QCD simulation is an equilibrium-state simulation, the rotating matter cannot be generated.
To overcome this difficulty, we rotate the reference frame.
As shown in Fig.~\ref{fig1}, the rotating matter seems rest in a rotating frame.
Particles do not flow in the rotating frame.
This state is in equilibrium and can be simulated in lattice QCD.
The transformation of the reference frame makes the simulation of the rotating matter possible.
 
In this study, we formulate lattice QCD in rotating frames toward the first-principle simulation of the rotating QCD matter \cite{Yamamoto:2013zwa}.

\section{Rotating lattice action}
We consider the coordinate transformation from a rest frame to a rotating frame in the Euclidean space-time.
We choose the Cartesian coordinate $x^{\mu} = (x,y,z,\tau)$ and rotate the coordinate around the $z$ axis.
The metric tensor of the rotating frame is
\begin{equation}
g_{\mu\nu} =
\begin{pmatrix}
1 & 0 & 0 & y\Omega \\
0 & 1 & 0 & -x\Omega \\
0 & 0 & 1 & 0 \\
y\Omega & -x\Omega & 0 & 1+r^2\Omega^2
\end{pmatrix}
.
\label{eqg}
\end{equation}
We substitute this metric tensor into the QCD action in a general curved space-time.

The continuum gluon action is
\begin{equation}
\begin{split}
S_G =& \int d^4x \ \frac{1}{g_{\rm YM}^2} \tr [ (1+r^2\Omega^2) F_{xy}F_{xy} + (1+y^2\Omega^2) F_{xz}F_{xz} \\
& + (1+x^2\Omega^2) F_{yz}F_{yz} + F_{x\tau}F_{x\tau} + F_{y\tau}F_{y\tau} + F_{z\tau}F_{z\tau} \\
& + 2y\Omega F_{xy}F_{y\tau} - 2x\Omega F_{yx}F_{x\tau} + 2y\Omega F_{xz}F_{z\tau} - 2x\Omega F_{yz}F_{z\tau} + 2xy \Omega^2 F_{xz}F_{zy} ].
\end{split}
\end{equation}
We discretize this action on the lattice.
The gluon field strength is constructed from the gauge invariant loops of the link variables $U_\mu(x)$.
The squared terms, e.g., $F_{xy}F_{xy}$, are constructed from the`` plaquette.''
We take the clover-type average of four plaquettes as
\begin{equation}
\bar{U}_{\mu\nu} = \frac{1}{4} \left( \parbox[c]{65pt}{\includegraphics[scale=0.35]{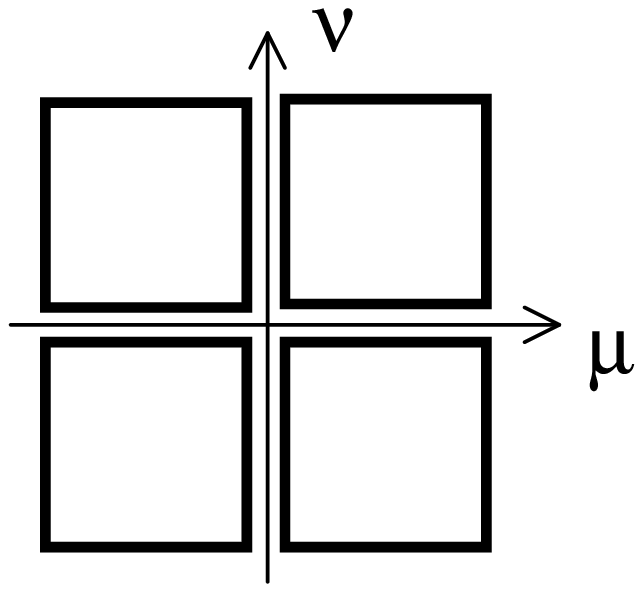}} \right).
\end{equation}
In the rotating frame, the gluon action includes the non-squared terms, e.g., $F_{xy}F_{y\tau}$, which break parity and time-reversal symmetry.
The non-squared terms are constructed from the ``chair-type'' loop \cite{Iwasaki:1983ck}.
We take the (anti-)symmetric average of eight chair-type loops as
\begin{equation}
\bar{V}_{\mu\nu\rho} = \frac{1}{8} \left( \parbox[c]{70pt}{\includegraphics[scale=0.35]{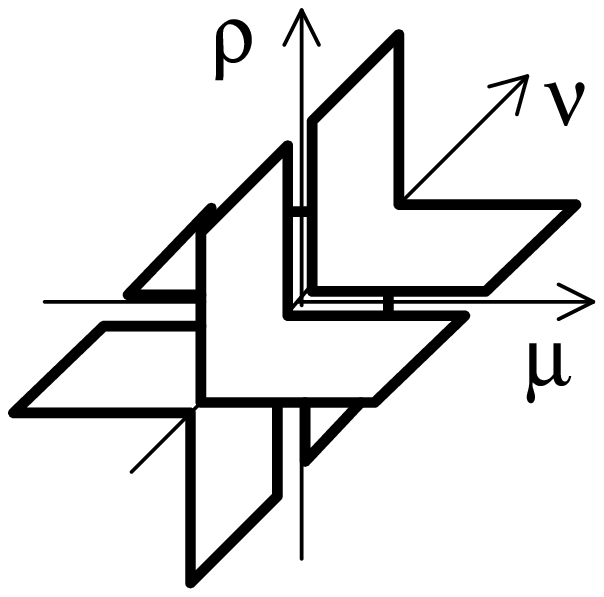}} - \parbox[c]{70pt}{\includegraphics[scale=0.35]{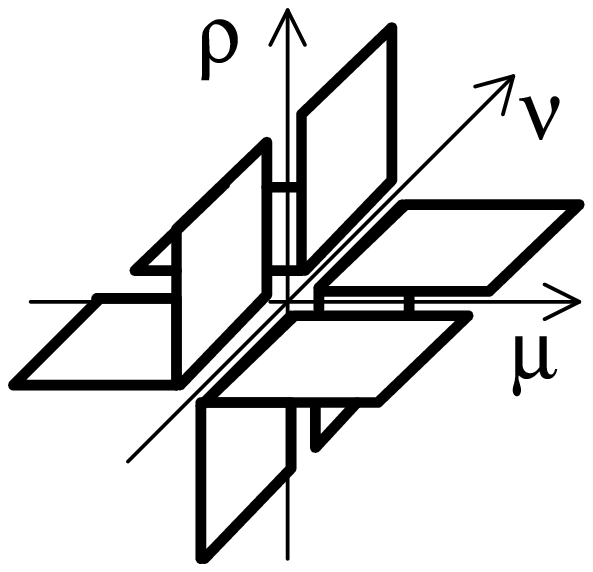}} \right).
\end{equation}
The lattice gauge action is
\begin{equation}
\begin{split}
S_G =& \sum_{x} \beta \bigg[ (1+r^2\Omega^2) \left(1-\frac{1}{N_c}\Retr \bar{U}_{xy} \right) + (1+y^2\Omega^2) \left(1-\frac{1}{N_c}\Retr \bar{U}_{xz} \right) \\
& + (1+x^2\Omega^2) \left(1-\frac{1}{N_c}\Retr \bar{U}_{yz} \right) + 3-\frac{1}{N_c}\Retr \left( \bar{U}_{x\tau} + \bar{U}_{y\tau} + \bar{U}_{z\tau} \right) \\
& -\frac{1}{N_c}\Retr \big( y\Omega \bar{V}_{xy\tau} - x\Omega \bar{V}_{yx\tau} + y\Omega \bar{V}_{xz\tau} - x\Omega \bar{V}_{yz\tau} + xy \Omega^2 \bar{V}_{xzy} \big) \bigg].
\end{split}
\end{equation}
The bare lattice coupling is $\beta = 2N_c/g_{\rm YM}^2$.

The continuum fermion action is
\begin{equation}
S_F = \int d^4x \ \bar{\psi} \bigg[ (\gamma^1 - y\Omega \gamma^4) D_x + (\gamma^2 + x\Omega \gamma^4) D_y + \gamma^3 D_z + \gamma^4 \left( D_\tau +i \Omega \frac{\sigma^{12}}{2} \right) +m \bigg] \psi.
\end{equation}
with $\sigma^{12} = \frac{i}{2}(\gamma^1\gamma^2 - \gamma^2\gamma^1)$.
As a result of rotation, the Dirac operator includes the orbit-rotation coupling term $\gamma^4 \Omega (xD_y-yD_x)$ and the spin-rotation coupling term $i \gamma^4 \Omega \sigma^{12}/2$.
We discretize this fermion action using the Wilson fermion as
\begin{equation}
\begin{split}
S_F =& \sum_{x_1,x_2} \bar{\psi}(x_1) \bigg[ \delta_{x_1,x_2} - \kappa \bigg\{(1-\gamma^1 + y\Omega \gamma^4) T_{x+} + (1+\gamma^1 - y\Omega \gamma^4) T_{x-} \\
& + (1-\gamma^2 - x\Omega \gamma^4) T_{y+} + (1+\gamma^2 + x\Omega \gamma^4) T_{y-} + (1-\gamma^3) T_{z+} + (1+\gamma^3) T_{z -} \\
& + (1-\gamma^4) \exp \left(i a\Omega \frac{\sigma^{12}}{2} \right) T_{\tau +} + (1+\gamma^4) \exp \left(-i a\Omega \frac{\sigma^{12}}{2} \right) T_{\tau -} \bigg\} \bigg] \psi(x_2)
\end{split}
\end{equation}
with $T_{\mu +} \equiv U_{\mu}(x_1)\delta_{x_1+\hat{\mu},x_2}$ and $T_{\mu -} \equiv U^\dagger_{\mu}(x_2)\delta_{x_1-\hat{\mu},x_2}$.
The bare hopping parameter is $\kappa = 1/(2am+8)$.

There are two kinds of the rotation in the Euclidean space-time: the ``Euclidean'' rotation and the ``Minkowskian'' rotation.
The order of two operations, the Wick rotation $\tau = -it$ and the spatial rotation $\theta = \theta_{\rm rest}-\Omega t$, is essential.
The Minkowskian rotation $\Omega = -\partial \theta / \partial t$ is defined as the spatial rotation before the Wick rotation, and the Euclidean rotation $\Omega = -\partial \theta / \partial \tau$ is defined as the spatial rotation after the Wick rotation.
As long as the analytic continuation to the original Minkowski space-time is validated, these rotations produce the same end result.
In the above equations, we adopted the Euclidean rotation.
For the Minkowskian rotation, the angular velocity is replaced as $\Omega \to i\Omega$.
Both of the gluon action and the fermion action becomes complex.
Thus, there is the sign problem in the Minkowskian rotation.
On the other hand, there is no sign problem in the Euclidean rotation.
(This is similar to the case of external electric fields. There is the sign problem in the Minkowskian electric field $E_j = \partial A_j / \partial t - \partial A_0 / \partial x^j$ and no sign problem in the Euclidean electric field $E_j = \partial A_j / \partial \tau - \partial A_4 / \partial x^j$ \cite{Yamamoto:2012bd}.)

\section{Numerical test}

We performed the quenched SU(3) Monte Carlo simulation.
The lattice size is $N_x \times N_y \times N_z \times N_\tau=13 \times 13 \times 12 \times 12$.
We set the bare lattice coupling $\beta = 5.9$ and the bare hopping parameter is $\kappa = 0.1583$, where the lattice spacing is $a\simeq 0.10$ fm and the meson mass ratio is $m_\pi/m_\rho\simeq 0.59$ \cite{Aoki:2002fd}.
We adopted the Euclidean rotation and restricted the angular velocity $\Omega$ only to small values.
The lattice is rotated around the $z$ axis.
In the $x$ and $y$ directions, we take the Dirichlet boundary conditions.
The range of the $x$-$y$ plane is $x=[-6a,6a]$ and $y=[-6a,6a]$.
In the $z$ and $\tau$ directions, we take boundary conditions in the same manner as the usual lattice simulation.

As the first observable, we calculate angular momentum.
Before discussing the simulation result, let us recall the case of a classical point particle.
In a rest frame, the classical Lagrangian is $\mathcal{L} = mr^2\dot{\theta}^2_{\rm rest}/2$ and the classical solution is $J_{\rm clas}=0$.
In a rotating frame, the classical Lagrangian is $\mathcal{L} = mr^2(\dot{\theta}+\Omega)^2/2$, and the classical solution is
\begin{equation}
J_{\rm clas}= -I\Omega =-mr^2\Omega,
\end{equation}
i.e., a finite angular momentum.
This is because the rest particle seems oppositely rotating.
The coefficient $I$ is the moment of inertia, and $I = mr^2$ for a point particle.
Using the lattice simulation in the rotating frame, we demonstrate that the same happens in QCD.

\begin{figure}[p]
\begin{center}
\includegraphics[scale=1.5]{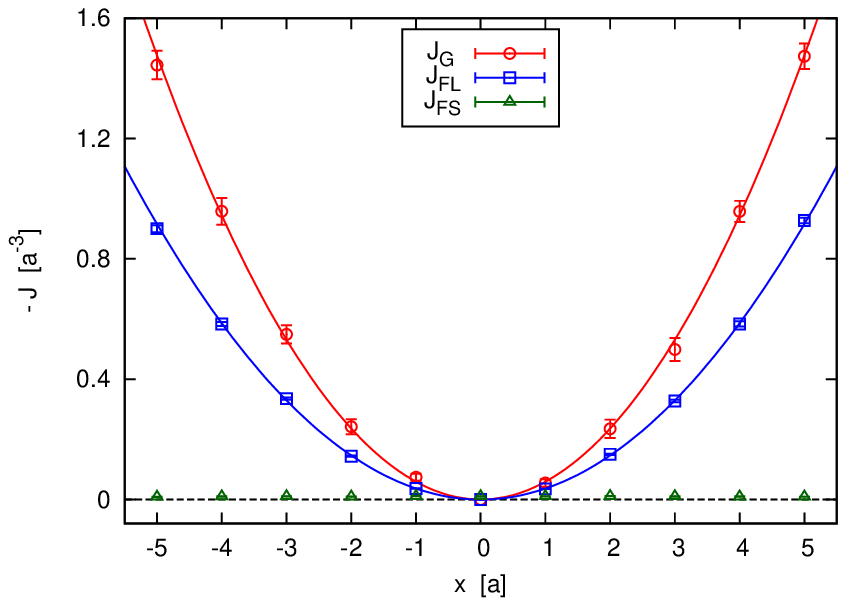}
\caption{\label{fig2}
Angular momentum density $J$ along the $x$ axis with the angular velocity $a\Omega = 0.06$.
The solid curves are quadratic fitting functions.
}
\end{center}

\begin{center}
\includegraphics[scale=1.5]{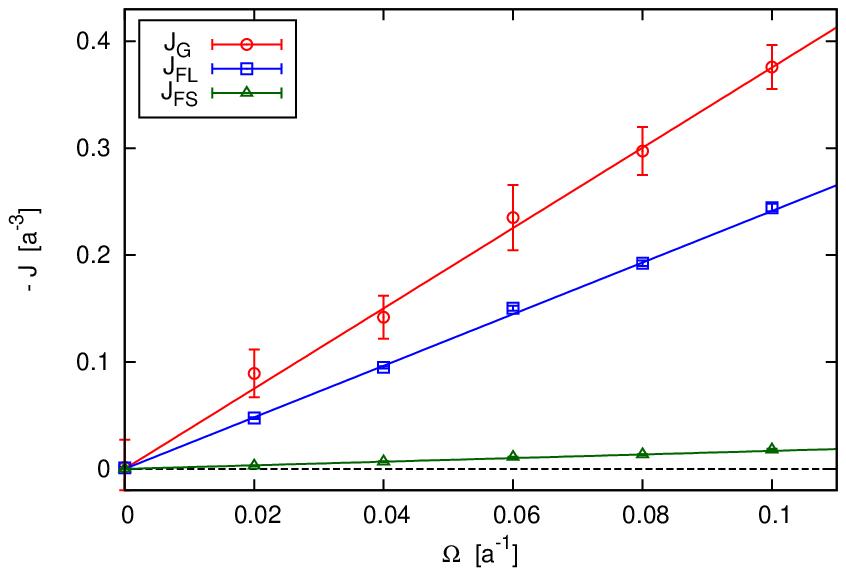}
\caption{\label{fig3}
Angular momentum density $J$ at $(x,y)=(2a,0)$ as a function of the angular velocity $\Omega$.
The solid curves are linear fitting functions.
}
\end{center}
\end{figure}

We analyzed the gluon angular momentum density
\begin{equation}
J_G = \bigg\langle \frac{1}{g_{\rm YM}^2} \tr ( 2y F_{xy}F_{y\tau} - 2x F_{yx}F_{x\tau} + 2y F_{xz}F_{z\tau} - 2x F_{yz}F_{z\tau} ) \bigg\rangle ,
\end{equation}
the fermion orbital angular momentum density
\begin{equation}
J_{FL} = \left\langle \bar{\psi} \gamma^4 (xD_y-yD_x) \psi \right\rangle,
\end{equation}
and the fermion spin angular momentum density
\begin{equation}
J_{FS} = \left\langle i \bar{\psi} \gamma^4 \frac{\sigma^{12}}{2} \psi \right\rangle.
\end{equation}
We discretize these operators in the same way as the lattice actions.
In Fig.~\ref{fig2}, we show the angular momentum density along the $x$ axis ($y=0$).
The angular velocity is fixed at $a\Omega = 0.06$.
$J_G$ and $J_{FL}$ are quadratic functions of the distance from the rotation axis.
$J_{FS}$ is independent of the distance (it is small but nonzero in the figure).
In Fig.~\ref{fig3}, we show the angular momentum density as a function of the angular velocity $\Omega$.
The spatial coordinate is fixed at $(x,y)=(2a,0)$.
All the angular momentum densities are proportional to the angular velocity.
Thus, the functional forms and the numerical coefficients are
\begin{eqnarray}
J_{G}  &=& - (0.94 \pm 0.01) a^{-4} \times r^2\Omega, \\
J_{FL} &=& - (0.60 \pm 0.01) a^{-4} \times r^2\Omega, \\
J_{FS} &=& - (0.17 \pm 0.01) a^{-2} \times \Omega.
\end{eqnarray}
$J_G$ and $J_{FL}$ are proportional to $r^2\Omega$, and thus behave like a classical point particle. 
$J_{FS}$ is independent of $r$ because spin is an intrinsic angular momentum.
Although the simulation is a fully quantum one, the qualitative behavior can be interpreted by an intuitive classical picture.
The numerical coefficients of $J_G$ and $J_{FL}$ are the inertial mass densities, and that of $J_{FS}$ is the density of the moment of inertia.

\section{Discussion}
We formulated lattice QCD in rotating frames.
In the above analysis of the angular momenta, the angular momenta seem finite in the rotating frame but the matter does not rotate in the rest frame.
The simulation of an actually rotating matter will be possible by introducing additional element in the rotating frame.
For example, if we introduce the Wilson line in the rotating frame, it corresponds to a rotating heavy-quark trajectory in the rest frame.
If we introduce an external anisotropic potential, we can generate the rotating matter with finite mass, which has been done in condensed matter physics for the rotating Bose-Einstein condensation \cite{Tsubota:2002}.
Once the rotating QCD matter is generated, there are many possible applications, e.g.,
rotating hadrons, the chiral vortical effect in rotating heavy-ion collisions \cite{Son:2009tf}, and quantum vortex nucleation in the core of compact stars \cite{Forbes:2001gj}.
More generally, this kind of the simulation can be applied to other field theories in curved space-time.

\section*{Acknowledgments}

A.~Y.~is supported by the Special Postdoctoral Research Program of RIKEN.
Y.~H.~is supported by the Japan Society for the Promotion of Science for Young Scientists and by JSPS Strategic Young Researcher Overseas Visits Program for Accelerating Brain Circulation.
The numerical simulations were performed by using the RIKEN Integrated Cluster of Clusters (RICC) facility.

\end{document}